\documentclass{aa}  
 \usepackage[varg]{txfonts}

\usepackage[switch]{lineno}
\usepackage{graphicx}
\usepackage{natbib,twoopt}
\usepackage[breaklinks=true, colorlinks=true, citecolor=blue, urlcolor=black]{hyperref} 
\bibpunct{(}{)}{;}{a}{}{,}             
\makeatletter
  \newcommandtwoopt{\citeads}[3][][]{\href{http://adsabs.harvard.edu/abs/#3}%
    {\def\hyper@linkstart##1##2{}%
     \let\hyper@linkend\@empty\citealp[#1][#2]{#3}}}
  \newcommandtwoopt{\citepads}[3][][]{\href{http://adsabs.harvard.edu/abs/#3}%
    {\def\hyper@linkstart##1##2{}%
     \let\hyper@linkend\@empty\citep[#1][#2]{#3}}}
  \newcommandtwoopt{\citetads}[3][][]{\href{http://adsabs.harvard.edu/abs/#3}%
    {\def\hyper@linkstart##1##2{}%
     \let\hyper@linkend\@empty\citet[#1][#2]{#3}}}
  \newcommandtwoopt{\citeyearads}[3][][]%
    {\href{http://adsabs.harvard.edu/abs/#3}
    {\def\hyper@linkstart##1##2{}%
     \let\hyper@linkend\@empty\citeyear[#1][#2]{#3}}}
\makeatother

\begin{document} 

   \title{Uncertainty estimation for time series classification}
   \subtitle{Exploring predictive uncertainty in transformer-based models for variable stars}

   \author{
          M. Cádiz-Leyton\inst{1, 2}\textsuperscript{\tiny\thanks{e-mail: \href{mailto:mcadiz2018@inf.udec.cl}{\texttt{mcadiz2018@inf.udec.cl}}}}
          \and 
          G. Cabrera-Vives\inst{1,3,4,5}\textsuperscript{\tiny\thanks{e-mail: \href{mailto:mcadiz2018@inf.udec.cl}{\texttt{guillecabrera@inf.udec.cl}}}}
          \and 
          P. Protopapas\inst{2}
          \and 
          D. Moreno-Cartagena\inst{1,2,3}
          \and 
          C. Donoso-Oliva\inst{3,5}
          \and
          I. Becker\inst{4,5,6}
          }

   \institute{
   Department of Computer Science, Universidad de Concepción, Edmundo Larenas 219, Concepción, Chile
    \and
    John A. Paulson School of Engineering and Applied Science, Harvard University, Cambridge, MA, 02138
   \and
    Center for Data and Artificial Intelligence, Universidad de Concepción, Edmundo Larenas 310, Concepción, Chile
    \and
    Millennium Institute of Astrophysics (MAS), Nuncio Monseñor Sotero Sanz 100, Of. 104, Providencia, Santiago, Chile
    \and
    Millennium Nucleus on Young Exoplanets and their Moons (YEMS), Chile
    \and
    Department of Computer Science, Pontificia Universidad Catolica de Chile, Macul, Santiago 7820436, Chile
    }


 
  \abstract
   {Classifying variable stars is key for understanding stellar evolution and galactic dynamics. With the demands of large astronomical surveys, machine learning models, especially attention-based neural networks, have become the state-of-the-art. While achieving high accuracy is crucial, enhancing model interpretability and uncertainty estimation is equally important to ensure that insights are both reliable and comprehensible.} 
   {We aim to enhance transformer-based models for classifying astronomical light curves by incorporating uncertainty estimation techniques to detect misclassified instances. We tested our methods on labeled datasets from MACHO, OGLE-III, and ATLAS, introducing a framework that significantly improves the reliability of automated classification for the next-generation surveys.}
   {We used Astromer, a transformer-based encoder designed for capturing representations of single-band light curves. We enhanced its capabilities by applying three methods for quantifying uncertainty: Monte Carlo Dropout (MC Dropout), Hierarchical Stochastic Attention (HSA), and a novel hybrid method combining both approaches, which we have named Hierarchical Attention with Monte Carlo Dropout (HA-MC Dropout). We compared these methods against a baseline of deep ensembles (DEs). To estimate uncertainty estimation scores for the misclassification task, we selected Sampled Maximum Probability (SMP), Probability Variance (PV), and Bayesian Active Learning by Disagreement (BALD) as uncertainty estimates.}
   {In predictive performance tests, HA-MC Dropout outperforms the baseline, achieving macro F1-scores of $79.8 \pm 0.5$ on OGLE, $84 \pm 1.3$ on ATLAS, and $76.6 \pm 1.8$ on MACHO. When comparing the PV score values, the quality of uncertainty estimation by HA-MC Dropout surpasses that of all other methods, with improvements of $2.5\pm 2.3$ for MACHO, $3.3\pm 2.1$ for ATLAS and $8.5\pm 1.6$ for OGLE-III.
   }
   {}

   \maketitle
%
\section{Introduction}

    The identification and classification of variable stars is crucial in advancing our understanding of the cosmos. For example, Cepheids and RR Lyrae stars are key rungs on the cosmological distance ladder \citep{feast2014cepheid, ngeow2015application}. The emergence of next-generation survey telescopes, such as the Vera Rubin Observatory and its Legacy Survey of Space and Time (\citealt{ivezic2019lsst}), presents new opportunities for analyzing an abundance of photometric observations. Such comprehensive data, with their corresponding time-stamps (i.e., light curves), are instrumental in detecting new classes of variable stars and uncovering previously unknown astronomical phenomena (\citealt{bassi2021classification}).    
    

    Classifying light curves is essential for analyzing variable stars; however, this task presents significant challenges due to heteroskedasticity, sparsity, and observational gaps \citep{mahabal2017deep}. Astronomy has transitioned from traditional feature-based analysis to advanced data-driven models enabled by deep learning \citep{smith2023astronomia}. This evolution is evidenced by the adoption of diverse architectures ranging from multilayer perceptrons \citep{karpenka2013simple} to Recurrent Neural Networks (RNNs) and Convolutional Neural Networks (CNNs)
    \citep{cabrerareyes, mahabal2017deep, protopapas2017recurrent, naul2018recurrent, carrasco2019deep, becker2020scalable, donoso2021effect}. 
    
    Building upon this evolution, self-attention-based models are setting new benchmarks \citep{donoso2022astromer, moreno2023positional, pan2023astroconformer, leung2023towards, parker2024AstroCLIP}. These models address applications from inferring black hole properties \citep{park2021inferring}, to denoising light curves \citep{morvan2022don} and specialized multi-band light-curve classification \citep{pimentel2022deep, cabrera2024atat}.
    
    As the complexity of models escalates, ensuring the reliability of classification results becomes increasingly critical. Within the framework of deep neural networks, uncertainty estimation is pivotal, as it not only enhances the confidence in predictions but also proves particularly beneficial in the detection of misclassifications, where uncertain predictions can signal potential errors \citep{gawlikowski2023survey}. Traditional techniques such as Bayesian Neural Networks (BNNs; \citealt{blundell2015weight}), which inherently model uncertainty by approximating a posterior distribution over model parameters, have been effective in astronomy \citep{moller2020supernnova, killestein2021transient,ciucua2021unveiling}.  However, their training stage requires high computational capacity, leading to increased resource use and extended convergence times.
    
    In this work, we present a methodology that combines deep attention-based classifiers with uncertainty estimation techniques for misclassification detection. Our model is based on Astromer, the transformer-based embedding approach proposed by \cite{donoso2022astromer}. For our analysis, we implemented three state-of-the-art uncertainty estimation approaches: deep ensembles (DE; \citealt{valdenegro2019deep, ganaie2022ensemble}, Monte Carlo Dropout (MC Dropout; \citealt{gal2016dropout}), and Hierarchical Stochastic Attention (HSA; \citealt{Pei_Wang_Szarvas_2022}). 
    We additionally propose an alternative approach, Hierarchical Attention with MC Dropout (HA-MC Dropout), which integrates a hierarchical attention structure with dropout activation during the inference stage.
    
    Although DEs are computationally expensive in practice, they establish a robust baseline for capturing uncertainty and facilitate comparisons with other techniques \citep{lakshminarayanan2017simple}. Thus, we used DEs as a baseline to compare the performance of MC Dropout, HSA and HA-MC Dropout within a variable star transformer-based classifier. 
    We further assess the uncertainty estimation capabilities of our models through statistical significance tests. Our main contributions are:
      
\begin{enumerate} 
    \item We present a methodology that fuses deep attention-based classifiers for astronomical time series with uncertainty estimation techniques.



    \item We establish through empirical evaluations that the HA-MC Dropout approach outperforms DEs in the context of variable star classification, emphasizing its utility as a primary technique for this application.

\end{enumerate}

This work is organized as follows: Sect. \ref{sec:methods} details the attention-based architecture used, outlines the methods employed, and formally defines the evaluation techniques used in our research. Section \ref{section:experimental} presents the  experimental setup, including dataset descriptions and the generation of misclassification instances. Our findings are then introduced in Sect. \ref{sec:results}, and the paper concludes with Sect. \ref{sec:discussionandconclusions}, which summarizes the key outcomes and insights derived from our work.

\section{Methods}
\label{sec:methods}

This section presents the methodology adopted in our work, focusing on the Astromer transformer-based model and its integration of uncertainty estimation techniques. We detail the four principal approaches: deep ensembles, Monte Carlo Dropout, hierarchical stochastic attention and the hierarchical attention with MC dropout to assess predictive uncertainty. Additionally, we explain the three types of uncertainty estimates employed for the misclassification task and outline the methodology for quality evaluation.
\subsection{Transformer-based model for light curves}

Our work uses an architecture inspired by Astromer, which is an encoder-decoder model derived from the principles of the Bidirectional Encoder Representations from Transformers (BERT; \citealt{devlin2019google}). Unlike BERT, which is specifically designed for natural language processing (NLP) tasks, Astromer is adapted for capturing embedding representations of astronomical light curves, serving as an automatic feature extractor for these time series.

The model includes several key components: positional encoding (PE), two self-attention blocks, and a long-short-term memory network classifier (LSTMs; \citealt{hochreiter1997long}). It processes single-band time series, each consisting of $L=200$ observations. These observations are characterized by a vector of magnitudes, $\vec{x} \in \mathbb{R}^L$, and corresponding timestamps, $\vec{t} \in \mathbb{R}^L$. The PE component encodes each timestamp using trigonometric functions, similar to those defined in \cite{vaswani2017attention}; however, it operates in the time domain of the light curve, rather than using index positions. Simultaneously, the magnitude data is fed into a feedforward (FF) neural network. For each point in the light curve, both the PE and the FF neural network output a vector of dimensionality $d_x = 256$. The outputs from the PE and the magnitudes FF neural network are added to produce $\vec{X} \in \mathbb{R}^{L \times d_x}$. This $200 \times 256$ matrix serves as the input representation for the standard transformer self-attention mechanism.

Transformers are composed of multiple layers of self-attention heads. A self-attention head transforms $\vec{X}$ into query, key, and value matrices ($\vec{Q}$, $\vec{K}$, $\vec{V}$) by using trainable weights $\vec{W_Q}, \vec{W_K}, \vec{W_V}\in \mathbb{R}^{d_x \times d_k}$:
\begin{equation}
\vec{Q} = \vec{X W_Q}, \quad \vec{K = XW_K}, \quad \vec{V = XW_V},
\end{equation}
where $d_k$ 
is the dimension of the output of the attention head for each observation. 

Attention weights $\alpha_{i,j}$ represent how much token $i$ attends token $j$. These weights are determined by calculating the dot product between the query and key vectors, $\vec{q}_i = \vec{W_Q}^\top\vec{x}_i$ and $\vec{k}_j= \vec{W_K}^\top\vec{x}_j$, and passing them through a softmax function in order to normalize them. Here, $\vec{x}_i$ represents the $i$-th row of $\vec{X}$. To stabilize the gradients during training, the dot product is divided by the square root of the dimension of the key vectors $\sqrt{d}$. In other words,
\begin{eqnarray}
\vec{\alpha}_{i} &=& \text{softmax}\left(\frac{\vec{q}_i^{\top} \vec{k}_1}{\sqrt{d_k}}, \frac{\vec{q}_i^{\top} \vec{k}_2}{\sqrt{d_k}}, \cdots, \frac{\vec{q}_i^{\top} \vec{k}_L}{\sqrt{d_k}}\right),\nonumber \\
&=& \text{softmax}\left( \vec{K}\vec{q}_i \right).
\label{eq:attscores}
\end{eqnarray}

The output $\vec{h}_i$ for each head is computed by applying the normalized attention weights to the corresponding values $\vec{v}_i=\vec{W_V}^\top\vec{x}_i$:
\begin{equation}
\vec{h}_i = \sum_j \alpha_{i,j} \vec{v}_j.
\label{eq:hvalues}
\end{equation}
In other words, the output of each head is the sum of the value vector representation of the input weighted by the attention payed to each of these vectors. This is usually written as
\begin{equation}
\mathrm{Attention}(\vec{Q}, \vec{K}, \vec{V}) = \text{softmax}\left(\frac{\vec{Q} \vec{K}^{\top}}{\sqrt{d_k}}\right)\vec{V},
\label{eq:hvalues}
\end{equation}
where the softmax is applied row-wise.

After calculating the outputs \(\vec{h_i}\) of each head, they are concatenated \([\vec{h_1}, \ldots, \vec{h}_{\text{\#heads}}]\) to form the final output of the multi-head attention block. Astromer has two blocks, each consisting of four heads with 64 neurons. To mitigate overfitting, it incorporates five dropout layers with a rate of 0.1. The output from this encoder is then fed into a two-layer LSTM network, which uses a softmax activation funcxtion to perform classification.

We initialized our encoders with pre-trained weights provided by the authors, who pretrained Astromer using 1,529,386 R-band light curves from the Massive Compact Halo Object survey (MACHO; \citealt{alcock2000macho}). Along with the classifier, we apply uncertainty quantification techniques to the encoder to enhance reliability in variable star classification. The model variants are further explained in subsequent sections.

\subsection{Deep Ensembles}
\label{subsec:deep_ens}

Using BNNs require significant computational resources, primarily due to the complex tuning necessary to achieve a consistent learning progress. In contrast, \cite{lakshminarayanan2017simple} presented deep ensembles as a scalable and viable non-Bayesian alternative. 

Consider a training dataset $\mathcal{D}=\{(\vec{x_i}, y_i)\}_{i=1}^N$, where each $\vec{x_i}\in \mathbb{R}^D$ is a $D$-dimensional feature vector. For a classification task, the labels $y_i$ are assumed to be one of $K$ classes, i.e., $y_i\in\{1, \ldots, K\}$. Now, consider a predictive model (such as a neural network) that, for a given input $\vec{x}$ outputs a prediction $\hat{y} = f_\theta(\vec{x})$, where $\vec{\theta}$ are the parameters of the model. When estimating uncertainties for a new data point $\vec{x}^*$, we aim at modeling the predictive distribution as:  
\begin{equation}
    p(y^*|\vec{x}^*, \mathcal{D}) = \int{p(y^*|\vec{x}^*, \vec{\theta})p(\vec{\theta}|\mathcal{D})d\theta}. \label{eq:predictive_distribution}
\end{equation}
This integral is often intractable and approximate inference is typically applied. Hence, one alternative for uncertainty estimation is the use of deep ensembles. In our work, the strategy used involves training $T$ independent neural network models, each with parameters $\{\vec{\theta}_t\}_{t=1}^T$. The predictive outcome for a new test point $\vec{x^*}$ is estimated by averaging the outputs from all $T$ models:
\begin{equation}
y^*\approx \frac{1}{T}\sum_{t=1}^Tf_{\theta_t}(\vec{x}^*). \label{eq:MC}
\end{equation}
Although this method is computationally intensive, it is easier to tune and can yield good performance while capturing predictive uncertainty.

We use DEs as the baseline uncertainty estimation technique. To calculate statistics over our uncertainty estimates, we trained ten ensembles, each consisting of ten deterministic models. These models were independently trained with different random seeds, which are used to initialize model parameters to different starting values. Additionally, we trained each model with different training chunks of the complete dataset. This approach introduces variability in the learning process, promoting diverse model behaviors and mitigating overfitting, thereby enhancing the robustness of our uncertainty estimates. The procedure was conducted for each survey.


\subsection{Monte Carlo Dropout (MC Dropout)}
\label{subsec:mcdp}

Dropout is a regularization technique used to prevent overfitting by deactivating neurons during the training stage \citep{srivastava2014dropout}. While initially aimed at mitigating overfitting, \cite{gal2016dropout} provided a theoretical framework for its application in uncertainty quantification. They demonstrated that Monte Carlo Dropout works as an approximation of Bayesian inference in deep Gaussian processes. This is achieved by sampling neuron activations from a Bernoulli distribution across all hidden layers of the neural network during both training and inference stages.

MC Dropout uses dropout during inference to approximate the predictive distribution of Eq.~\ref{eq:predictive_distribution}. This adaptation allows MC Dropout to mimic an ensemble of diverse models. Each stochastic pass temporarily deactivates a random subset of neurons, and the expected value of the output \( y^* \) for a given input \(\vec{x}^* \) is calculated by averaging the outputs from the model \( f_{\theta_t}(\vec{x}^*) \) over all samples, as shown in Eq.~\ref{eq:MC}. 

Consider the neural network layer $i$, which receives the output vector $\vec{x}_{i-1}$ from the preceding layer as its input. When dropout is applied with probability $p$, the output of layer $i$ can be defined as:
\begin{equation}
\vec{x}_i=  \sigma(\vec{x}_{i-1}|\vec{W_i},\vec{M_i}), 
\label{eq:MCDdropout}
\end{equation}
where $\sigma$ represents the activation function, $\vec{W}_i$ denotes the weights of layer $i$, and $\vec{M}_i$ are the variational parameters that modulate these weights during training and inference. The weights $\vec{W}_i$ are given by:
\begin{equation}
\vec{W}_i= {\vec{M}}_i \cdot \text{diag}([{z}_{i,j}]_{j=1}^{K_i}) , \quad\quad {z}_{i,j} \thicksim  Bernoulli(1 - p_i),
\label{eq:MCDdropoutWGHTS}
\end{equation}
here $z_{i,j}=0$ indicates that the neuron \(j\) in layer \(i - 1\) is dropped as an input to layer \(i\). 

This method offers advantages over DEs as it avoids the need to train multiple models while retaining the ability to estimate uncertainties by using the variability of subnetworks within a single model. The authors proposed applying this method to all hidden neural network layers. In adapting MC Dropout for transformers, following the methodology described by \cite{shelmanov2021certain}, we applied dropout to all layers in the model, including those within the attention blocks. This strategy captures uncertainty at multiple levels of abstraction by introducing stochasticity throughout the entire network, ensuring that diverse model behaviors are accounted for, leading to more robust uncertainty estimates than if dropout were applied only to the output layer.

\subsection{Hierarchical Stochastic Attention (HSA)}
\label{subsec:hsa}

\cite{Pei_Wang_Szarvas_2022} proposed Hierarchical Stochastic Attention, an approach for producing probabilistic outputs rather than deterministic ones by injecting stochasticity through the Gumbel-softmax distribution \citep{jang2016categorical}.  This uncertainty estimation method was applied to NLP tasks, achieving competitive predictive performance while allowing for uncertainty estimation. HSA is composed of two hierarchical stochastic self-attention mechanisms: stochasticity over the self-attention heads and over a set of learnable centroids.

Stochastic self-attention replaces the traditional softmax activation function of the attention heads with the Gumbel-softmax distribution, which approximates samples from a categorical distribution with class probabilities $\vec{\theta} = (\theta_1, \theta_2, \ldots, \theta_K)$ and associated logits values $\log(\theta_i)$. The Gumbel-softmax distribution considers a temperature $\tau$, and samples $\tilde{\vec{y}}\sim \mathcal{G}(\vec{\theta}, \tau)$ are computed as 
\begin{eqnarray}
\tilde{y}_{i} &=& \frac{\exp\left(\left(\log(\theta_i) + g_i\right) / \tau\right)}{\sum_{j=1}^K \exp\left(\left(\log(\theta_j) + g_j\right) / \tau\right)},
\label{eq:gumbel-softmax}
\end{eqnarray}
where $g_{i}$ represents i.i.d. samples from the Gumbel distribution, defined as $g_{i} = -\log(-\log(u))$, where $u$ is uniformly drawn from the interval [0,1]. This approximation facilitates discrete choice sampling and gradient-based optimization needed on neural networks and that can not be performed using a categorical distribution. Furthermore, using the temperature $\tau$, the Gumbel-softmax distribution can be smoothly annealed into a categorical distribution: as $\tau\to 0$, the Gumbel-softmax samples are identical to the categorical distribution.

In stochastic self-attention, logits are computed as the logarithm of the dot product between the query and the key vectors, and the attention weights are calculated during a forward pass as follows:
\begin{eqnarray}
\hat{\alpha}_{i,j} &=& \frac{\exp\left(\left(\log(\vec{q}_i^\top \vec{k}_{j}) + g_{i,j}\right) / \tau\right)}{\sum_{l=1}^L \exp\left(\left(\log(\vec{q}_i^\top \vec{k}_{l}^\top) + g_{i, l}\right) / \tau\right)},\\
\hat{\vec{\alpha}}_i&\sim& \mathcal{G}\left(\vec{K}\vec{q}_i, \tau\right),
\label{eq:stochastic_attention}
\end{eqnarray}
where $g_{i, j}$ represents i.i.d. samples from the Gumbel distribution. Here, the Gumbel-softmax distribution has a similar role to the traditional softmax by normalizing scores across all keys $k$, but allows to sample from it and estimate gradients efficiently.

The second component of HSA forces heads to pay stochastic attention to a set of centroids. Instead of directly attending to each key, HSA employs the Gumbel-softmax distribution to group keys around $c$ learnable centroids \(\vec{C} \in \mathbb{R}^{d_k \times c}\), where each centroid $\vec{c}_j$ represents the $j$-th column of $\vec{C}$ and matches the dimension of each key head. The model starts by stochastically paying attention to the centroids, and, then, a new set of keys $\{\tilde{\vec{k}}_i\}_{i=1}^L$ are calculated by weighting each centroid by this attention:
\begin{eqnarray}
\tilde{\vec{\alpha}}_i&\sim& \mathcal{G}\left(\vec{C}^\top\vec{k}_i, \tau\right),\\
\tilde{\vec{k}}_i &=& \sum_j \tilde{\alpha}_{i,j} \vec{c}_j.
\label{eq:centroid_attention}
\end{eqnarray}
Here, $g_{i,j}$ are again sampled from the Gumbel-softmax distribution. The new keys of Eq. \ref{eq:centroid_attention} are used to calculate the stochastic attention weights of Equation \ref{eq:stochastic_attention} which are used to combine the values $\{\vec{v}_i\}_{i=1}^L$ as:
\begin{eqnarray}
\hat{\vec{\alpha}}_i&\sim& \mathcal{G}\left(\tilde{\vec{K}}\vec{q}_i, \tau\right),\\
\vec{h}_i &=& \sum_{j=1}^L\hat{\alpha}_{i,j}\vec{v}_j,
\end{eqnarray}
where $\tilde{\vec{k}}_i$ is represented as the $i$-th row of matrix $\tilde{\vec{K}}\in \mathbb{R}^{L\times d_k}$.
Notice the hierarchical nature of HSA: the method employs the Gumbel-softmax distribution twice; initially to generate the new keys $\{\tilde{\vec{k}}_i\}_{i=1}^L$, and a second time to compute the outputs for each attention block $\{\vec{h}_i\}_{i=1}^L$.



We implemented the HSA approach in the Astromer attention mechanism, which captures predictive uncertainties through multiple forward passes during the inference stage, similar to the MC Dropout approach.

\subsection{Hierarchical Attention with MC Dropout (HA-MC)}
\label{subsec:hsa}
We propose a combination of MC Dropout and the Hierarchical Attention strategy for quantifying uncertainty in Astromer. This approach incorporates learnable centroids $\vec{c}_i$ in the attention blocks without applying the Gumbel-softmax distribution, but instead adding MC Dropout to the attention blocks. This allows us to achieve a regularization effect through the key head’s attention to the centroids,
while adding stochasticity without the use of the Gumbel-softmax distribution. The approach is defined as follows: 
\begin{eqnarray}
\tilde{\vec{\alpha}}_{i} &=& \text{softmax}\left(\frac{\vec{C}^{\top} \vec{k}_i}{\sqrt{d_k}}\right),\\
\tilde{\vec{k}}_i &=& \sum_j \tilde{\alpha}_{i,j} \vec{c}_j,
\label{eq:attscores}
\end{eqnarray}
where \(d_k\) is the original scaling factor, and $\vec{C}$ is the matrix with the centroids as columns from Eq. \ref{eq:centroid_attention}. The final attention scores are then computed using:
\begin{eqnarray}
\tilde{\vec{\alpha}}_{i} &=& \text{softmax}\left(\frac{\tilde{\vec{K}} \vec{q}_i}{\sqrt{d_k}}\right),\\
\vec{h}_i &=& \sum_j \hat{\alpha}_{i,j} \vec{v}_j,
\label{eq:attscores}
\end{eqnarray}
In this context, stochasticity is introduced by activating the dropout layers within each attention block during the inference stage.



\subsection{Uncertainty Estimates} 

When a model is trained using only a maximum likelihood approach, the softmax activation function is prone to generate overconfident predictions \citep{guo2017calibration}. To explore and quantify the uncertainty inherent in our model's predictions, we employ uncertainty estimates (UEs). These estimates are not designed to assess the quality of the uncertainty quantification but to measure the extent and variability of uncertainty itself. 

For the MC Dropout model variants, each UE is calculated by conducting $T$ forward pass inference runs with dropout activated. In the case of HSA, the variation in the $T$ inference runs is obtained through samples drawn from the Gumbel-softmax distribution. For the deep ensembles baseline, $T$ corresponds to the number of independent models. Using these $T$ inference runs, we calculate the following uncertainty estimates:



\begin{itemize}
    \item Sampled Maximum Probability (\(\mathrm{SMP}\)):
        \begin{equation}
            1 - \max_{c \in C } \overline{p}(y=c|x),
        \end{equation}
        where  $\overline{p}(y=c|x) = \frac{1}{T} \sum_{t=1}^{T} p_{t}(y=c|x)$ is the average probability of each class \(c\) across \(T\) forward passes for a given input \(x\) (being \(p_t(y=c|x)\) the probability of class \(c\) at the \(t\)-th forward pass inference run). The SMP provides an intuitive measure of confidence by considering the maximum mean predicted probability across classes. 

    \item Probability Variance (\(\mathrm{PV}\); \citealt{gal2017deep}):
        \begin{equation}
           \frac{1}{C} \sum_{c=1}^{C}  \left( \frac{1}{T}  \sum_{t=1}^{T} \left( p_{t}(y=c|x) -  \overline{p}(y=c|x)\right)^2 \right).
            \label{eq:pv}
            \end{equation}
        This is the variance averaged over all \(C\) classes. PV assesses how consistently the model predicts the same class probabilities across different inference runs, providing an insight into the predictive stability.

    \item Bayesian Active Learning by Disagreement (\(\mathrm{BALD}\); \citealt{houlsby2011bayesian}):
        \begin{equation}
            -\sum_{c=1}^{C}\overline{p^{c}}\log(\overline{p^{c}}) + \frac{1}{T}\sum_{t=1}^{T}\sum_{c=1}^{C}  p_{t}^{c} \log (p_{t}^{c}),
        \end{equation}
        where \(\overline{p^{c}} = \overline{p}(y=c|x)\), and \(p_{t}^{c}= p_{t}(y=c|x)\). The first term is the entropy of the average predictions across the ensemble of models, whereas the second term calculates the mean entropy of individual predictions from each model within the ensemble, reflecting the average uncertainty inherent in each model's predictions. \(\mathrm{BALD}\), unlike the other two uncertainty estimates, is entropy-based which can be interpreted as a measure of total uncertainty \citep{depeweg2018decomposition,NEURIPS2018_3ea2db50}. 
\end{itemize}

\subsection{Evaluating uncertainty estimation via misclassification detection} \label{sec:missclassification}

A well-calibrated uncertainty estimation model should exhibit high uncertainty for incorrect predictions while maintaining low uncertainty for correct ones. To evaluate the performance of uncertainty estimation, we define a misclassification detection task. Although all of our models were trained as multiclass classifiers, we reformulate the evaluation during the testing phase as a binary classification problem focused on identifying misclassifications. This approach is aligned with the work done by \cite{shelmanov2021certain} and \cite{vazhentsev2022uncertainty} for the NLP tasks. 

Specifically, we construct new binary instances $ \tilde e_i$ as follows:
    \begin{equation}
     {\tilde{e}_i} =
    \begin{cases}
        1, &  y_i \neq  \hat{y}_i, \\
        0, &  y_i =  \hat{y}_i, 
    \end{cases}
    \end{equation}
where $y_i$ is the true label, and $\hat{y}_i$ is the original predicted label. The new instances $\tilde{e}_i$ indicate whether the model made a mistake in predicting the label of the variable source. 

To measure the quality of UE, we compute the Receiver Operating Characteristic Area Under the Curve (ROC AUC) scores based on the binary labels $\tilde {e}_i$ and their corresponding UE scores. We use the UEs values as discrimination values to build the ROC curve, which plots the true positive rate (TPR) against the false positive rate (FPR) across various thresholds \citep{swets1988measuring}. For a specific UE threshold, the TPR and FPR are calculated as 
\[ \mathrm{TPR} = \frac{\mathrm{TP}}{\mathrm{TP} + \mathrm{FN}}, \]
and
\[ \mathrm{FPR} = \frac{\mathrm{FP}}{\mathrm{FP} + \mathrm{TN}}, \]
where: 
\begin{itemize}
    \item TP are the number of instances that have an UE higher than the threslhold, and $\tilde{e}_i = 1$ (missclassified instances with high uncertainty),
    \item TN are the number of instances that have an UE lower than the threslhold, and $\tilde{e}_i = 0$ (correctly classified instances with low uncertainty),
    \item FP are the number of instances that have an UE higher than the threslhold, and $\tilde{e}_i = 0$ (correctly classified instances with high uncertainty),
    \item FN are the number of instances that have an UE lower than the threslhold, and $\tilde{e}_i = 1$ (missclassified instances with low uncertainty).
\end{itemize}

We evaluated statistical significance of our results using the Wilcoxon-Mund Whitney (WMW) test. The WMW test, a non-parametric alternative to the \( t \)-test, evaluates data based on ranks rather than assuming normality or equal variances. This test ranks all observations from two independent samples, \( X \) and \( Y \), and calculates the test statistic, \( U \), using the ranks from the smaller sample. The null hypothesis states that \( X \) and \( Y \) have identical distributions. Significant deviations in \( U \) suggests differing distributions, with statistical significance indicating disparities in central tendencies \citep{fay2010wilcoxon, pett2015nonparametric}. 

\section{Experimental Setup} 
In this section, we detail our experimental setup, where we evaluate the models using three labeled astronomical catalogs. 
We balanced the number of samples per class in small chunks for both the training and testing stages. 
The experiments were conducted on a Nvidia RTX A5000 GPU, our focus was on detecting misclassifications by converting the multiclass task into a binary problem, thus providing a clearly understanding of model performance under varied data scenarios.

\label{section:experimental}
\subsection{Data and training}
\label{subsec:datadesc}

In this study, we compared the performance of baseline and the proposed models on three labeled catalogs of variable stars: the Optical Gravitational Lensing Experiment (OGLE-III; \citealt{udalski2004optical}), the Asteroid Terrestrial-impact Last Alert System (ATLAS; \citealt{heinze2018first}), and MACHO dataset. We considered the classification scheme and filtering methods previously selected by \cite{becker_2020} and utilized by \cite{donoso2022astromer}, as detailed in Table \ref{tab:data_classes}. These catalogs contain light curves observed through different spectral filters, offering a broad spectrum of data for analysis.

Preserving in-domain integrity during model testing was crucial to our methodology. This principle required evaluating models on data with distributional characteristics similar to those of the training set. Therefore, we avoided combining the test sets from the three catalogs during inference, despite some catalogs sharing classes. This approach highlights the importance of testing models in conditions that mirror their training environment. Although the OGLE-III and MACHO datasets share similar wavelength ranges, the distinct spectral band of the ATLAS catalog indicates the need for a meticulous in-domain evaluation approach.

\begin{table}[h]
    \caption{Variable stars classes  of each survey associated to the corresponding tag.}
    \centering
    \renewcommand{\arraystretch}{1.35}
    \begin{tabular}{ccc}
    \hline\hline

    Dataset & Tag & Class\\
   
    \hline
     & Cep 0 & Cepheid type I \\
     & Cep 1 & Cepheid type II\\
     MACHO$^{(1)}$  & EC & Eclipsing binary \\
     & LVP &  Long period variable\\
        & RRab & RR Lyrae type ab \\
        & RRc & RR Lyrae type c \\
        \hline
     & CB & Close Binaries\\
     & DB & Detached Binary\\
     ATLAS$^{(2)}$ & Mira & Mira\\
     & Pulse & RR Lyrae, $\delta$-Scuti, Cepheids \\
        \hline
     & EC  & Eclipsing binary\\
     & ED & Detached Binary\\
      & ESD &  Semi-detached Binary \\
     & Mira &Mira \\
        & OSARG & Small-amplitude red giant\\
    OGLE-III$^{(3)}$    & RRab &  RRLyra type ab\\
        & RRc &  RRLyra type c\\
        & SRV & Semi-regular variable \\
        & Cep & Cepheid\\
        & DSct & Delta Scuti\\
    \hline
    
    \label{tab:data_classes}
    \end{tabular}
    \vskip -0.2in
    \tablebib{Classification scheme used by \citet{donoso2022astromer} for the following datasets, listed in the order of their appearance: (1) \citet{alcock2000macho}; (2) \citet{heinze2018first}; (3) \citet{udalski2004optical}.}
\end{table}

To emulate scenarios with a small amount of data, we selected 500 samples per class for training and 100 samples per class for test sets from the raw data (see Table \ref{tab:data_training}). A validation set was created by randomly selecting 30\% of the training set.

\begin{table}[!htb]
    \caption{Data distribution in terms of the number of light curves.}
    \renewcommand{\arraystretch}{1.35}
    \centering
    \begin{tabular}{cccc}
    \hline\hline
Dataset &  Raw data & Training set &  Test set\\
    \hline
    MACHO & 21\,444 & 3\,000 & 600\\
    ATLAS & 4\,719\,921 & 2\,000 & 400\\
    OGLE-III & 393\,103 & 5\,000 & 1\,000\\
    \hline
    \label{tab:data_training}
    \end{tabular}
    \vskip -0.25in
\end{table}


%

We used ten ensembles for the baseline and ten variants models per approach. A a single test set per survey was used to compare the performance of the models. Consequently, we collected ten predictions for each approach, enabling us to calculate the mean and standard deviation of the samples to conduct significance testing.
    
For the optimization technique, we chose Adam \citep{kingma2014adam} with a learning rate of $10^{-3}$. The batch size was of 512, and as a regularization technique, we used Early Stopping with a patience of 20 epochs on the validation loss. We used this same hyperparameter settings for each experiment.

\section{Results} 
\label{sec:results}

\subsection{Predictive performance}

\begin{table}[ht]
\caption{Summary of macro average multiclass metrics scores (\%) on MACHO, ATLAS and OGLE-III test sets.}
\label{tab:multiclass_performance}
\centering
\small
\renewcommand{\arraystretch}{1.35}
\begin{tabular}{c c c c c} 
\hline\hline
Method & Metric & MACHO & ATLAS & OGLE-III \\ 
\hline
 & F1  & 68.6{$\pm$}1.7 & 77.8{$\pm$}2.6 & 67.3{$\pm$}2.9 \\
{Baseline}   & Accuracy  & 69.5{$\pm$}1.6 & 77.7{$\pm$}2.5 & 68.5{$\pm$}2.9 \\
                          & Precision  & 69.3{$\pm$}1.7&78.1{$\pm$}2.6&68.6{$\pm$}2.8\\
\hline
 & F1 & 68.0{$\pm$}0.6 & 78.8{$\pm$}0.7 & 66.8{$\pm$}1.6 \\
 
{MC Dropout} & Accuracy  & 69.3{$\pm$}0.5 & 78.8{$\pm$}0.7 & 68.4{$\pm$}1.4 \\                          
                          & Precision  & 68.4{$\pm$}0.6&79.3{$\pm$}0.8&67.4{$\pm$}1.6\\
\hline
                & F1 & 74.6{$\pm$}1.8 & 82.1{$\pm$}2.2 & 75.7{$\pm$}1.9 \\
 {HSA}   & Accuracy  & 75.3{$\pm$}1.8 & 82.2{$\pm$}2.1 & 76.7{$\pm$}1.7 \\
                  & Precision  & 75.3{$\pm$}1.6&82.9{$\pm$}1.8&77.3{$\pm$}1.3\\
\hline
 & F1 & \textbf{76.6{$\pm$}0.8}& \textbf{84.0{$\pm$}1.3 }& \textbf{79.8{$\pm$}0.5}\\
{HA-MC Dropout}         & Accuracy  & \textbf{77.5{$\pm$}0.7} & \textbf{84.0{$\pm$}1.3} & \textbf{80.5{$\pm$}0.4} \\
        & Precision  &\textbf{77.1{$\pm$}0.5}&\textbf{84.3{$\pm$}1.2}&\textbf{80.0{$\pm$}0.3} \\
\hline
\end{tabular}
\vskip -0.10in
\end{table}

We evaluated the predictive performance using the macro-average of multiclass classification metrics over the three single-band datasets: MACHO, ATLAS and OGLE-III. Table \ref{tab:multiclass_performance} presents the test-sets F1 score, accuracy and precision of the HSA, MC Dropout, HA-MC Dropout and DEs methods.

The DEs baseline serves as a consistent benchmark, achieving macro F1 scores of 68.6/77.8/67.3 on MACHO, ATLAS, and OGLE-III, respectively. The MC Dropout method demonstrates a marginal performance improvement on the ATLAS test-set, with a similar macro F1-score and accuracy of 78.8\%, indicating limited gains over the baseline. 

In contrast, the HSA method significantly improves performance across all datasets, even when tested on datasets with varying numbers of classes (6 classes in MACHO, 4 in ATLAS, and 10 in OGLE-III). This demonstrates its ability to perform well across varying levels of class complexity, reflecting its capacity to capture more intricate patterns. However, it exhibits more variability in terms of standard deviation compared to other approaches, with the variability in performance metrics close to 2\%. This increased variability may be attributed to the stochasticity introduced by the Gumbel-softmax distribution, which can impact the predictive performance consistency across different runs. Consequently, while HSA improves overall performance, its stability across different runs may be less consistent than that of other approaches.

The proposed HA-MC Dropout significantly outperforms the other methods achieving higher scores across all metrics. HA-MC Dropout achieves macro F1-scores of 76.6/84.0/79.8 on MACHO, ATLAS, and OGLE-III, respectively, marking a substantial improvement in multiclass classification performance. Particularly in the dataset with 10 classes, OGLE-III, HA-MC Dropout obtained more than a 10\% improvement compared to the baseline, with f1-score/accuracy/precision of 79.8/80.5/80.0. These results indicate that the integration of hierarchical attention mechanisms with MC Dropout not only enhances predictive accuracy but also provides a more reliable model with reduced variance in performance metrics. This may be explained by the way stochasticity is injected into the model: unlike HSA that relies on the Gumbel-softmax distribution, HA-MC Dropout utilizes the activation of dropout in the inference stage to estimate uncertainty.

Summarizing, two of the methods, HSA and HA-MC Dropout, surpass the DEs baseline. This indicates that combining hierarchical attention with a stochastic component yields improvements in multiclass classification tasks. The comparison presented in Table \ref{tab:multiclass_performance} represents the first stage of our analysis, focusing on the multiclass evaluation to establish a general understanding of the performance enhancements provided by these methods. However, the interpretability and trustworthiness of the results are further supported by the uncertainty estimation, which provides insights into the reliability of our classification outcomes. We extend our analysis by considering the impact of uncertainty estimation on misclassification task in Sect. \ref{sec:mis_task_results}, providing a more comprehensive evaluation of the robustness and reliability of the different methods in handling complex classification tasks.

\subsection{Predictive uncertainty}
\label{sec:mis_task_results}

We evaluate the methods in terms of the predictive uncertainty using the misclassification detection task, as described in Sec.~\ref{sec:missclassification}, on the MACHO, ATLAS and OGLE-III test-sets. Table \ref{tab:roc_auc} shows the ROC AUC scores for the different UEs.  In this context, the ideal classifier is one that aligns uncertainty estimates with the misclassification task: misclassified instances should be associated with high uncertainty, while correctly classified instances should correspond to low uncertainty.
We compare the baseline DE method against MC Dropout, HSA, and HA-MC Dropout. The evaluation metric used is the absolute ROC AUC score, which quantifies the ability of the model to discern between missclassifications and correct classifications by using the UEs as discrimination scores. The uncertainty estimates used to calculate each ROC AUC are SMP, PV, and BALD. Specifically, for the baseline, the mean absolute ROC AUC is presented, whereas for the other methods, we report the performance differences relative to the baseline's corresponding uncertainty estimates, highlighting any statistically significant improvements ($p$-values $\leq$ 0.05). Note that the results are grouped by dataset.
Standard deviations are reported to reflect the variability across ten model iterations. For the baseline, this includes results from ten independent ensemble runs, where each run consists of ten separately trained deterministic models. For MC Dropout, HSA, and HA-MC Dropout, the results are based on ten independently trained models, with  $T=10$ stochastic inference runs performed per object for each model.

The DEs consistently achieve an average ROC AUC exceeding 70\% across all uncertainty estimates and datasets. This performance highlights their capability to identify potential errors through probabilistic outputs. MC Dropout excels in capturing predictive uncertainty when using PV and BALD scores for ROC AUC calculation, surpassing the baseline in misclassification detection tasks across all datasets. This is specially noticeable on the OGLE-III dataset, where the incremental percentage differences in ROC AUC relative to the baseline are $4.3\pm1.6\%$ and $4.2\pm1.6\%$, respectively.

\begin{table}[h]
\caption{Uncertainty estimates in the misclassification task on the MACHO, ATLAS and OGLE-III test sets.}
\label{tab:roc_auc}
\centering
\small
\renewcommand{\arraystretch}{1.35}
\begin{tabular}{c c c c c}
\hline\hline
Method        & UEs  & MACHO  & ATLAS   & OGLE-III    \\ \hline
        
& SMP  & 75.6{$\pm$}1.6&85.9{$\pm$}2.1&82.2{$\pm$}1.1\\
{Baseline} & PV   & 71.4{$\pm$}2.3&82.0{$\pm$}2.2&73.4{$\pm$}1.7\\
& BALD & 70.0{$\pm$}2.6&80.8{$\pm$}2.4&74.1{$\pm$}1.8\\ \hline                       
{\begin{tabular}[c]{@{}l@{}}\end{tabular}}  
& SMP  & 0.3{$\pm$}2.0&0.2{$\pm$}1.5&0.0{$\pm$}1.0  \\ 
MC Dropout& PV   &\textbf{1.4{$\pm$}2.4}&\textbf{1.5{$\pm$}2.3}&\textbf{4.3{$\pm$}1.6}\\ 
& BALD &\textbf{1.8{$\pm$}2.4}&\textbf{1.8{$\pm$}2.6}&\textbf{4.2{$\pm$}1.6}\\ 
\hline
{\begin{tabular}[c]{@{}l@{}}\end{tabular}}  
& SMP  & -1.5{$\pm$}1.8&-1.2{$\pm$}2.1&-0.1{$\pm$}1.0  \\ 
HSA & PV   &0.5{$\pm$}2.4&\textbf{2.6{$\pm$}2.2}&\textbf{5.9{$\pm$}1.5}\\ 
& BALD &-0.3{$\pm$}3.1&\textbf{2.3{$\pm$}2.5}&\textbf{2.9{$\pm$}1.3}\\ 
\hline
{\begin{tabular}[c]{@{}l@{}}\end{tabular}}  
& SMP  & 0.3{$\pm$}1.8&0.6{$\pm$}1.8&\textbf{2.1{$\pm$}1.2} \\ 
HA-MC Dropout& PV   &
\textbf{2.5{$\pm$}2.3} &\textbf{3.3{$\pm$}2.1} &\textbf{8.5{$\pm$}1.6}\\ 
& BALD & \textbf{2.1{$\pm$}2.7}& \textbf{3.8{$\pm$}2.4} & \textbf{6.9{$\pm$}1.6}\\ 
\hline
\end{tabular}

\tablefoot{The baseline presents the mean and standard deviation of the absolute ROC AUC scores (\%). For all other methods, the values indicate the difference in ROC AUC relative to the baseline, calculated for each uncertainty estimate. Statistically significant improvements ($p$-values $\leq$ 0.05) over the baseline for the corresponding UE are highlighted in bold font.}
\vskip -0.1in
\end{table}
Conversely, HSA registers significant improvements with PV and BALD on the ATLAS and OGLE-III datasets. However, it fails to demonstrate an enhancement on the MACHO test set with any of the uncertainty estimates provided. In the case of SMP scores, noticeable differences between the baseline arise where negative values indicate a reduction relative to the baseline.

\begin{figure*}[h]
\begin{center}
\includegraphics[ width=18cm ]{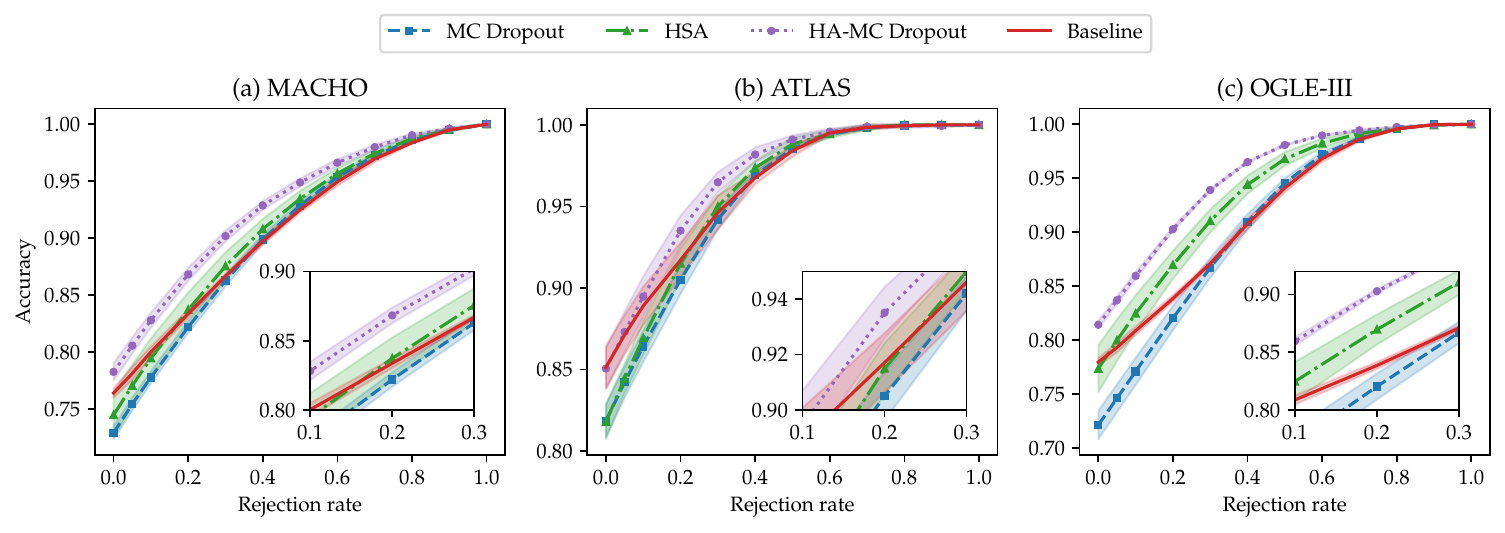}
\caption{Figures (a), (b), and (c) presents the accuracy-rejection curves for the MACHO, ATLAS, and OGLE-III datasets, respectively. Techniques compared include MC Dropout (dash-squared line), HSA (dash-triangular line), HA-MC Dropout (dash-dotted line), and the baseline model (solid line). Insets zoom into the lower rejection rate region (0-0.3) to emphasize differences at low rejection levels.}
\label{fig:acc}
\end{center}
\vskip -0.10in
\end{figure*}

To address these challenges, we introduced the HA-MC Dropout method, which, as aforementioned, combines the strengths of both MC Dropout and HSA. This hybrid approach significantly outperforms both individual methods, especially in SMP scores, where it rivals or even surpasses the baseline. For instance, it achieves an improvement of $2.1\pm1.2\%$ on MACHO. The most substantial improvements are observed with PV on OGLE-III, where it reaches $8.5\pm1.6\%$, doubling the improvement achieved by MC Dropout. Additionally, with PV, the improvement on MACHO is $2.5\pm2.3\%$, and using BALD yields $3.8\pm2.4\%$ on ATLAS. Consequently, HA-MC Dropout not only mitigates the limitations of its constituent methods but also establishes a new benchmark for balancing predictive accuracy and uncertainty quantification across diverse datasets.
\subsubsection{Accuracy-rejection plots}

We present a practical application of our misclassification framework by using the accuracy-rejection plots \citep{nadeem2009accuracy} for MACHO, ATLAS and OGLE-III test sets, as illustrated in Figures \ref{fig:acc} (a), (b) and (c). These plots emulate a scenario reflecting a hybrid machine-human behavior, wherein the machine abstains from classifying the most uncertain samples. This approach allows us to visualize accuracy as a function of the rejection rate. In all figures, confidence levels were assessed using the PV score, and the shaded areas represent the standard deviation across ten iterations of each approach. Additionally, each plot includes a zoomed inset focusing on the rejection rate interval from $0.1$ to $0.3$, providing a detailed comparison of the various methods within this specific range.

As a guideline, in Fig. \ref{fig:acc} (a), the accuracy-rejection plot for the MACHO test set suggests that maintaining an accuracy threshold above 80\% requires a rejection rate of $\sim 0.15$ for MC Dropout and $\sim 0.1$ for the DE baseline and HSA. HA-MC Dropout is able to keep a 80\% accuracy by rejecting less than the $\sim 5\%$ most uncertain predicted labels in the MACHO dataset. 
Notably, the HA-MC Dropout method surpasses both the baseline and other techniques for every rejection rate, highlighting its potential. Although HSA shows a marginally higher mean accuracy compared to the baseline, the baseline demonstrates lower variability at a rejection rate of 0.2, indicating higher consistency. Meanwhile, MC Dropout aligns closely with the baseline performance until a rejection rate of $\sim0.4$.

Figure \ref{fig:acc} (b) presents the accuracy-rejection plot for the ATLAS test set. Below a rejection rate of $\sim0.2$, HA-MC Dropout shows a performance comparable to the baseline. However, at a rejection rate of $\sim0.2$, the HA-MC Dropout method achieves a mean accuracy of approximately 0.93, surpassing the baseline by about 0.02 points, demonstrating its superior performance. 

Finally, Fig. \ref{fig:acc} (c) details the accuracy-rejection scenario for the OGLE-III dataset, indicating that both HA-MC Dropout and HSA outperform the baseline, with HA-MC Dropout needing a 20\% rejection rate to achieve a mean accuracy score of 0.90. HSA follows closely with an accuracy of 0.87, while the baseline achieves 0.84 and MC Dropout 0.82. This scenario underscores the necessity for expert intervention to maintain high accuracy levels, demonstrating that HA-MC Dropout achieves superior results with minimal expert involvement.

The analysis of these plots presents that, across all datasets evaluated, the performance curves of the MC Dropout approach align with the baseline model for a rejection rate higher than $\sim 0.4$.  This consistency at a specified threshold highlights the capability of the MC Dropout to maintain accuracy while also providing estimates of uncertainty. Conversely, the HA-MC Dropout method is a better option to other methods in all datasets. Despite the baseline showing lower standard deviation  in the OGLE and MACHO datasets, the overall accuracy score consistency across different astronomical datasets highlights the resilience of HA-MC Dropout. This robust performance affirms the cost-efficiency of implementing HA-MC Dropout for estimating uncertainty on transformer-based classifiers, making them well-suited for real-world applications.

\section{Discussion and Conclusions}
\label{sec:discussionandconclusions}
We have investigated the application of uncertainty estimation techniques to enhance the reliability and interpretability of transformer-based models for light curve classification in the context of variable star analysis. By implementing and evaluating deep ensembles, Monte Carlo Dropout, Hierarchical Stochastic Attention, and our proposed hybrid method, HA-MC Dropout in Astromer, we have demonstrated the potential of these techniques in capturing predictive uncertainty and improving misclassification detection.

Our empirical results highlight that HA-MC Dropout consistently outperforms other methods in terms of predictive accuracy and uncertainty estimation across various datasets. This suggests that integrating hierarchical attention mechanisms with Monte Carlo Dropout offers a powerful approach for enhancing the robustness and reliability of transformer-based models in complex classification tasks. The superior performance of HA-MC Dropout, particularly in scenarios with limited data, highlights its potential for real-world applications in scenarios with limited data and class distribution challenges.

The accuracy-rejection plots provide valuable insights into the practical implications of our work. This plots demonstrate that HA-MC Dropout enables the model to achieve higher accuracy levels with less rejected samples, showcasing its potential for automating the classification process while maintaining high confidence in the results. The consistent performance of MC Dropout across different datasets further reinforces its value as a viable alternative to the other approaches. Hence, it offers a computationally efficient and effective method for uncertainty estimation in transformer-based models.


The findings of this study have significant implications for the future of variable star classification, particularly in the era of next-generation large-scale astronomical surveys such as the LSST. The ability to quantify uncertainty and detect misclassifications will be crucial in ensuring the reliability and interpretability of automated classification systems. The work presented here offers a promising step towards achieving this goal, paving the way for more robust and trustworthy analysis of astronomical light curves.

Future research directions include exploring the application of our work into multi-band light curves model (e.g, \citealt{cabrera2024atat}). Additionally, considering the impact of different data preprocessing and augmentation strategies on uncertainty estimation could provide valuable insights into improving the performance of the transformer-based model in challenging scenarios. Human feedback for objects classified by the model with low certainty can also be added into a human-in-the-loop framework (see e.g. \citealt{richards2011active, masci2014automated, martinez2018high, ishida2019optimizing, kennamer2020active, leoni2022fink}).

Summarizing, HA-MC Dropout has proven to be competitive against the DEs baseline in three astronomical datasets with different variable star taxonomies. Transformer-based models have established their status as the state-of-the-art across various fields. We emphasize the significance of developing reliable models that can reduce computational expenses when being trained: DEs need to train multiple models, while the MC Dropout strategy uses a single trained model. We believe that the capacity to accurately assess uncertainty can economize human labor while also enhancing confidence in the conclusions derived from these models. 


\begin{acknowledgements}
The authors acknowledge support from the National Agency for Research and Development (ANID) grants: FONDECYT regular 1231877 (MCL, GCV, DMC, CDO); Millennium Science Initiative Program – NCN2021 080 (GCV, CDO)
and ICN12 009 (GCV, CDO, IB).
\end{acknowledgements}

\bibliographystyle{aa}
\bibliography{example}

\end{document}